\documentclass[twocolumn]{aastex631}

\newcommand{\Lagr}{$\mathcal{L}$}
\usepackage{amsmath}

\begin{document}


\title{PKS~2332$-$017 and PMN J1916$-$1519: Candidate Blazar Counterparts to 
Two High-energy Neutrino Events}

\author{Shunhao Ji}
\affiliation{Department of Astronomy, School of Physics and Astronomy, Yunnan
University, Kunming 650091, China; jishunhao@mail.ynu.edu.cn, wangzx20@ynu.edu.cn}

\author[0000-0003-1984-3852]{Zhongxiang Wang}
\affiliation{Department of Astronomy, School of Physics and Astronomy, Yunnan
University, Kunming 650091, China; jishunhao@mail.ynu.edu.cn, wangzx20@ynu.edu.cn}
\affiliation{Shanghai Astronomical Observatory, Chinese Academy of Sciences, 80
Nandan Road, Shanghai 200030, China}

\author{Dong Zheng}
\affiliation{Department of Astronomy, School of Physics and Astronomy, Yunnan
University, Kunming 650091, China; jishunhao@mail.ynu.edu.cn, wangzx20@ynu.edu.cn}

\begin{abstract}
We report our counterpart identification study for two high-energy neutrino 
events IC-130127A and IC-131204A listed in the IceCube Event Catalog 
of Alert Tracks. These two events belong to Gold alerts, which have
a significant probability of being of astrophysical origin.
	Within the events' 90\% positional uncertainty regions, we 
respectively find PKS~2332$-$017 and PMN J1916$-$1519. The first source is a 
flat-spectrum radio quasar at redshift $z= 1.18$ and the second a blazar of 
an uncertain type with photometric $z= 0.968$. As they correspondingly
had a $\gamma$-ray flare temporally coincident with the arrival times of 
IC-130127A and IC-131204A,
we identify them as the respective neutrino emitters. Detailed analysis of
the $\gamma$-ray data for the two blazars, obtained with the Large Area 
Telescope (LAT) onboard {\it the Fermi Gamma-ray Space Telescope (Fermi)}, 
is conducted.  The two flares respectively from PKS~2332$-$017 and 
PMN~J1916$-$1519 lasted $\sim$4\,yr and $\sim$4\,month, and showed possible 
emission hardening by containing high-energy $\sim$2--10\,GeV photons in
the emissions.
Accompanying the flare of PKS~2332$-$017, optical and MIR brightening variations
were also observed. We discuss the properties of the two sources and
	compare the properties with those of the previously reported 
(candidate) neutrino-emitting blazars.
\end{abstract}

\keywords{Blazars (164); Gamma-ray sources (633); Neutrino astronomy (1100)}

\section{Introduction} \label{sec:intro}

Neutrino astronomy has great potential of providing us with more and
different information for various celestial objects in the Universe, and 
the IceCube \citep{Aartsen+17}, a neutrino observatory built at the South 
Pole, is one of the large facilities leading the field in exploring the 
neutrino sky.
Since 2013, IceCube has been detecting high-energy neutrino events that
more likely have an astrophysical origin \citep{Aartsen+13}, and it was
in 2017 that the first convincing association was established. The case
was between a $\sim$0.3\,PeV neutrino event, named IC-170922A, and a flaring 
blazar TXS~0506+056, resulting from the multimessenger campaign
\citep{txs0506a,txs0506b}. Following this first association case, the neutrino 
emission from the nearby Seyfert galaxy NGC~1068 at a significance of 
4.2$\sigma$ and that from the Galactic plane at a 4.5$\sigma$ significance 
level have also been reportedly revealed \citep{ngc1068,gal}.

Blazars are a sub-type of Active Galactic Nuclei (AGNs) having their jets 
pointing close to our line of sight. They have been considered as possible 
sources of high-energy neutrinos (e.g., \citealt{mannheim,muc+03,Murase+18}). 
They are 
generally categorized as two types, Flat Spectrum Radio Quasars (FSRQs) and 
BL Lac objects (BL Lacs), based on the observed strengths of their optical 
emission lines \citep{urry+95,Scarpa+97}. Electrons accelerated
in the blazars' jets emit strongly via the synchrotron process in the radio to 
X-ray bands. Thus there are also three sub-classes used to describe their 
synchrotron emissions and spectral energy distributions (SEDs) based on 
the frequency values of the synchrotron 
peak $\nu^{\rm syn}_{\rm pk}$ \citep{Abdo+10}:
low synchrotron peaked (LSP), intermediate synchrotron 
peaked (ISP), and high synchrotron peaked (HSP). 
Protons should also be
accelerated in the jets, and the high-energy neutrino emission may originate 
from the photopion processes, as the result of the interaction between
relativistic protons and low-energy photons; charged and neutral pions are 
produced in the interaction, 
which subsequently decay and produce neutrinos and high-energy photons. 

\begin{figure*}[!ht]
\centering
\includegraphics[width=0.49\textwidth]{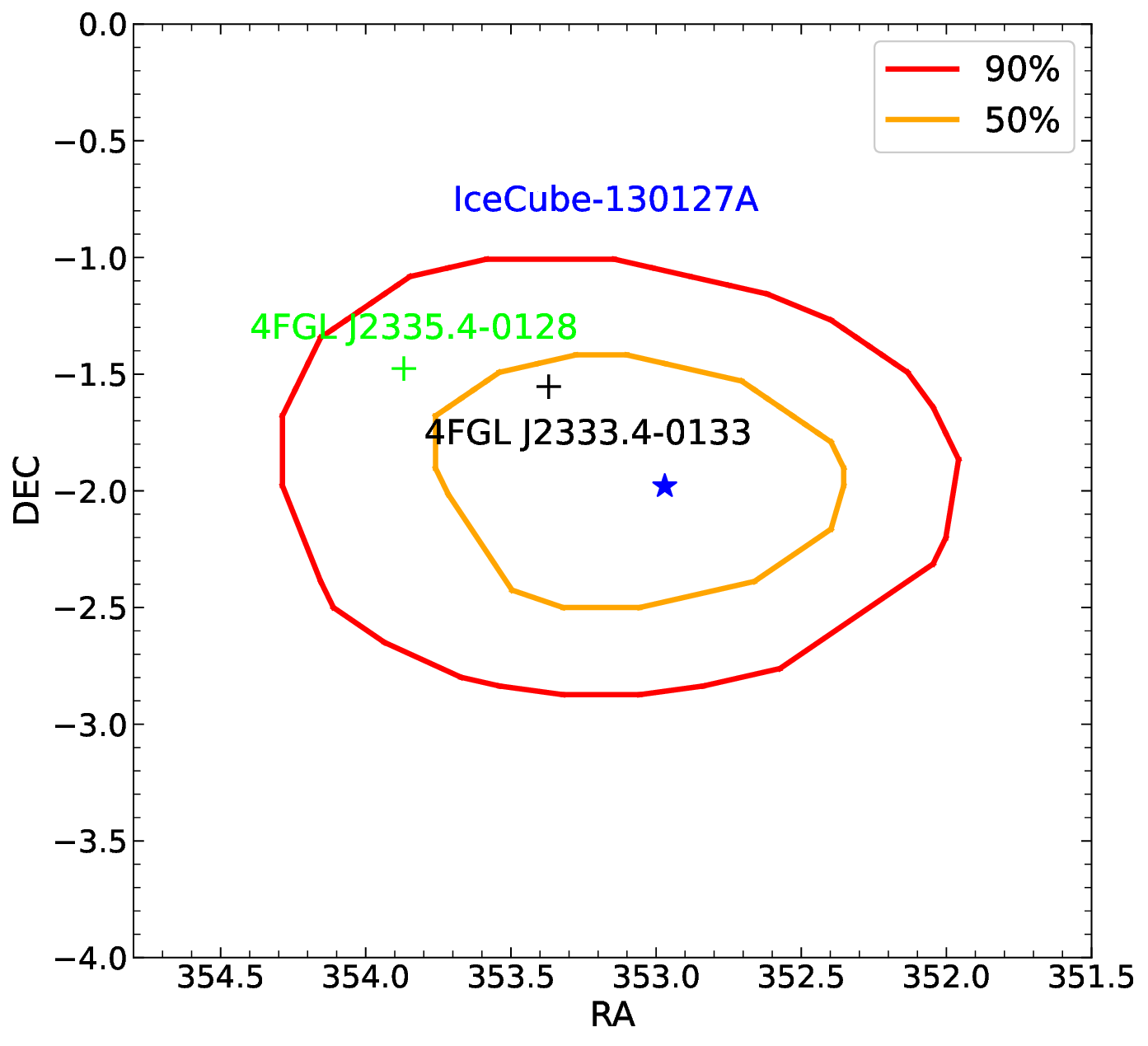}
\includegraphics[width=0.49\textwidth]{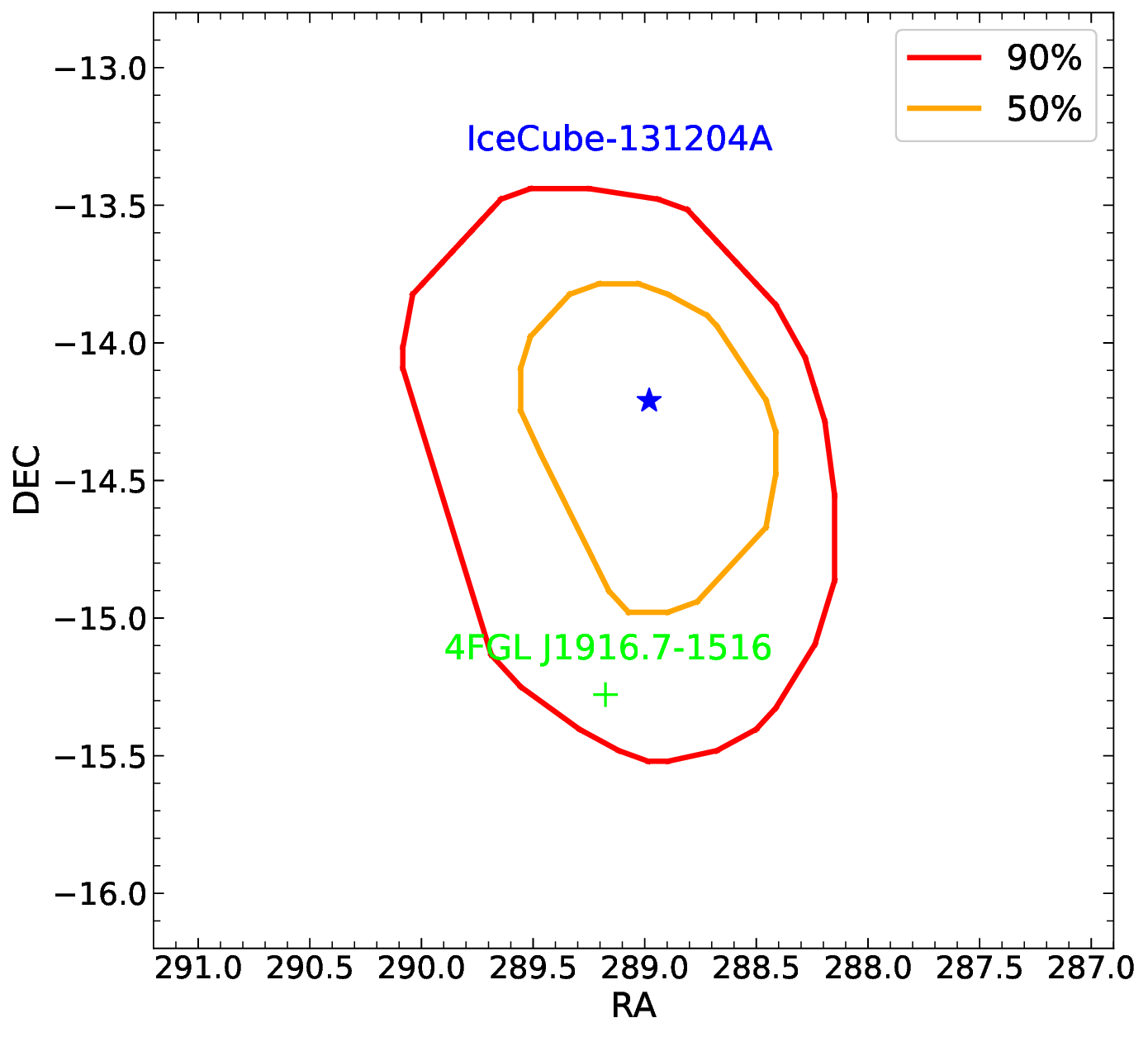}
\caption{Positional uncertainty regions (50\% and 90\%) of 
	IC-130127A ({\it left}) and IC-131204A ({\it right}), marked by
	the (respectively orange and red) contours. The green pluses mark 
	the positions of the neutrino-source candidates.
\label{fig:maps}}
\end{figure*}

Given the neutrino detection from TXS 0506+056, efforts have been made
to find more cases. A few blazars have been reported as the possible neutrino 
sources. For instance, there are two LSP BL Lacs, 
GB6 J1040+0617 \citep{gar+19} and 
PKS 2254+074 \citep{Ji+24}, two ISP BL Lacs, MG3 J225517+2409 \citep{fra+20} 
and PKS~0735+178 \citep{sah+23}, an HSP BL Lac, 
3HSP J095507.9+355101 \citep{gio+20}, and three LSP FSRQs, 
PKS B1424$-$418 \citep{kad+16}, GB6 J2113+1121 \citep{liao+22}, and 
NVSS J171822+423948 \citep{jiang+24}. The criteria for the identifications
of these blazar sources
follow the case of TXS 0506+056: each of them was positionally coincident 
with a high-energy neutrino event and at the arrival time of the neutrino, 
the blazar was having a $\gamma$-ray flare (or an X-ray flare
in the case of 3HSP J095507.9+355101).

Recently, the updated IceCube Event Catalog of Alert Tracks (IceCat-1) has been
released, which contains more than 100 neutrino track-like events 
likely of astrophysical origin (with energy $\geq$ 100\,TeV) detected from 
2011 to 2023 
\citep{Abbasi+23}. In IceCat-1, based on a quantity called 
``signalness'', the neutrino events having a significant 
probability of being an astrophysical neutrino are classified as Gold alerts, 
and they have an  
average signalness above 50\% \citep{Abbasi+23}. 
We went through the events in IceCat-1 and found two possible
association cases. The high-energy neutrino events are IC-130127A and
IC-131204A, both belonging to the Gold alerts. Based on the general 
criteria
currently considered in the association studies, the neutrinos possibly came 
respectively from two blazars, PKS~2332$-$017 and PMN~J1916$-$1519.
We report the association studies here.
Throughout this work, we used the following cosmological parameters, 
$H_0$ = 67.7 km s$^{-1}$ Mpc$^{-1}$, $\Omega_m$ = 0.32, and 
$\Omega_\Lambda$ = 0.68 \citep{Planck18}.

\section{Multiband and $\gamma$-ray Data}\label{data}

\subsection{Multiband Data}\label{optical}

Optical light-curve data were obtained from the Zwicky Transient Facility 
(ZTF; \citealt{Bellm+19}) and Catalina Real-Time Transient Survey (CRTS; 
\citealt{Drake+09}).
The bands of the data are ZTF $zg$ and $zr$, and CRTS $V$.
To ensure the good-quality of the ZTF data, we required 
catflags = 0 and $chi < 4$ for the downloaded magnitude data points. 

Mid-infrared (MIR) light-curve data were obtained from the
NEOWISE Single-exposure Source Database (\citealt{Mainzer+14}), where 
the bands are Wide-field Infrared Survey Explorer (WISE) 
W1 (3.4\,$\mu$m) and W2 (4.6\,$\mu$m).

We also searched the X-ray data for the sources. Unfortunately, there
were no observations from the X-ray Telescope (XRT; \citealt{Burrows+05}) 
onboard 
{\it the Neil Gehrels Swift Observatory}. For eROSITA \citep{Predehl+21}, 
both sources are not in the sky region of its recently released X-ray data.

\begin{figure*}[!ht]
\centering
\includegraphics[width=0.49\textwidth]{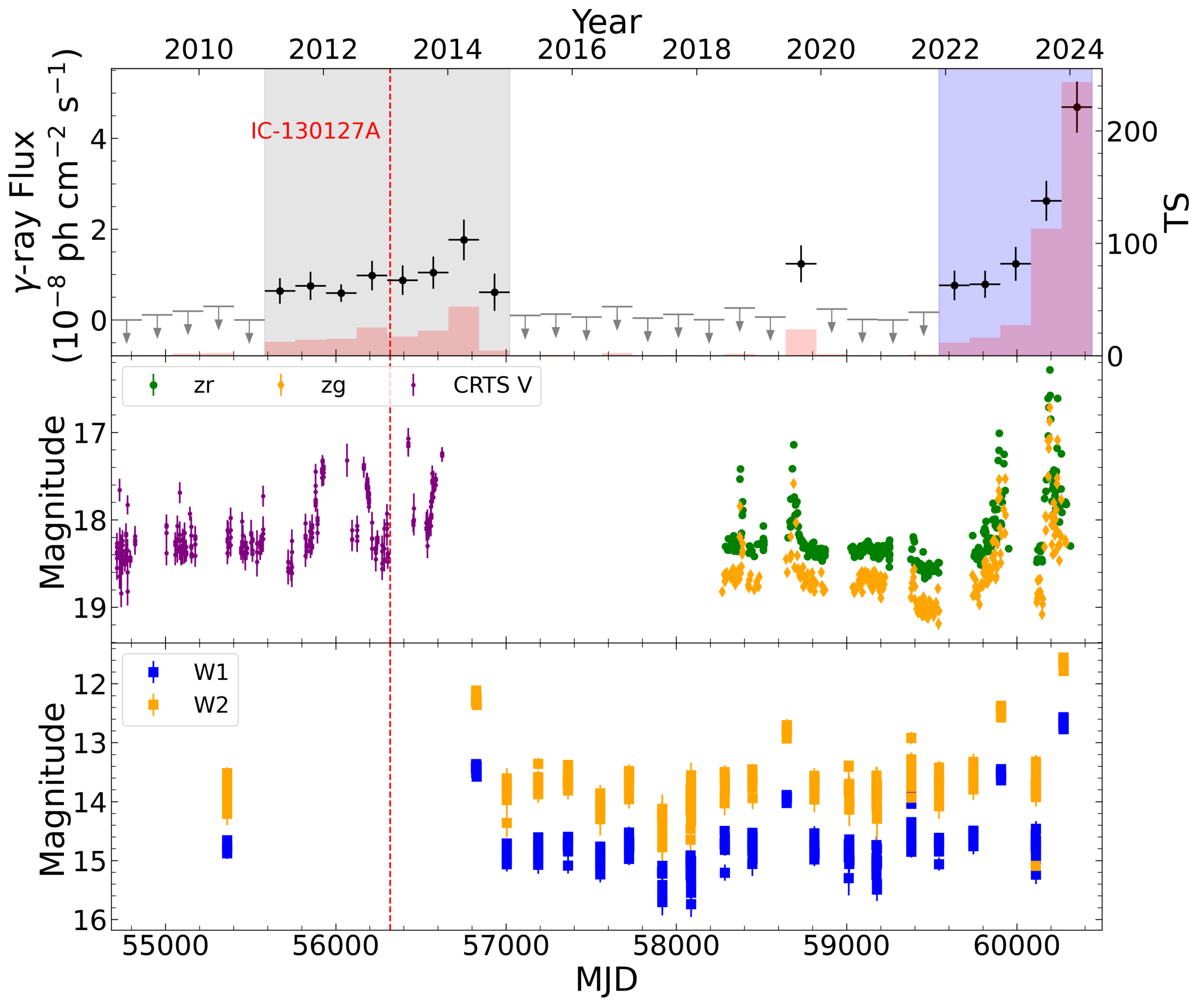}
\includegraphics[width=0.49\textwidth]{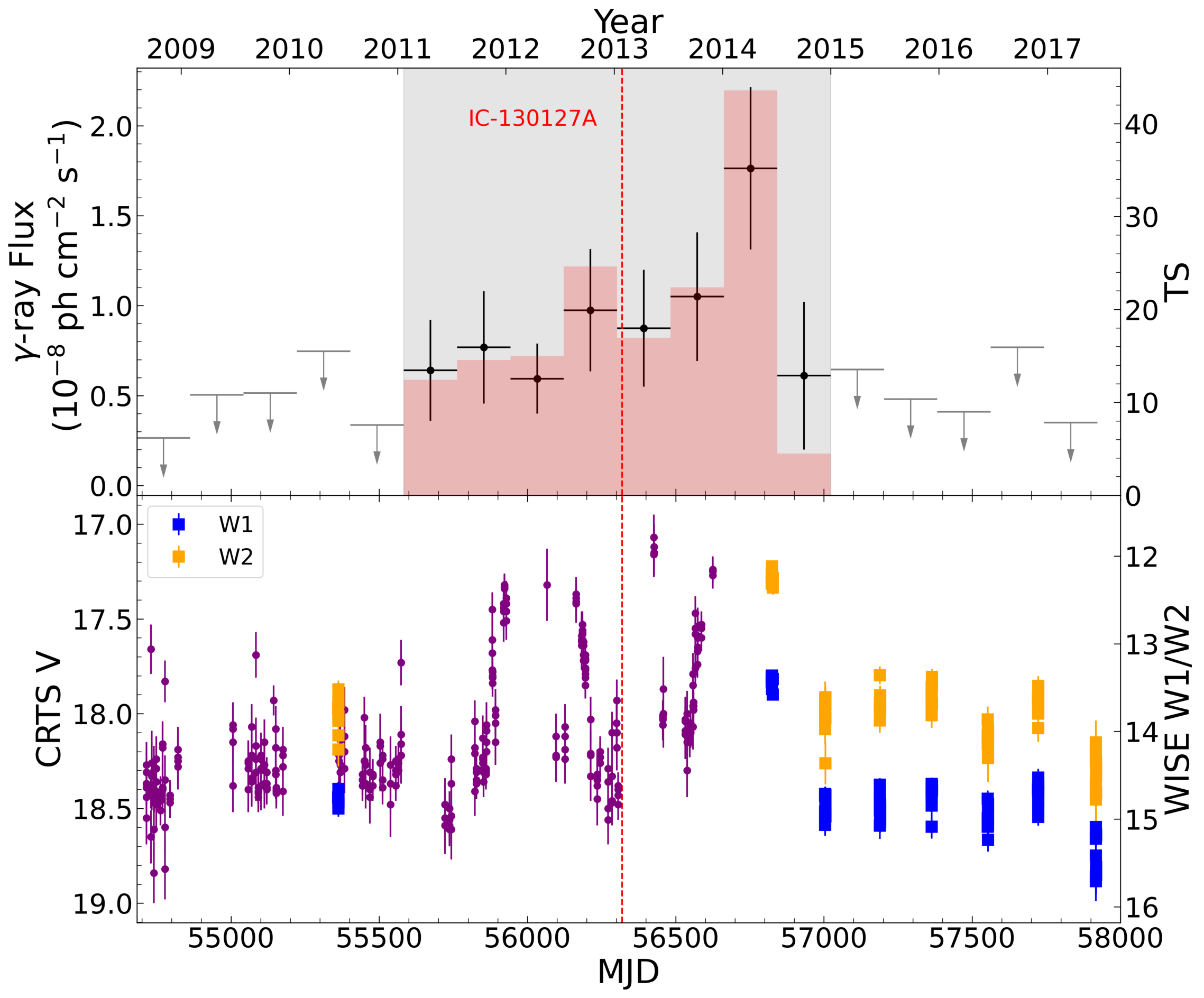}
\caption{{\it Left:} 180-day binned $\gamma$-ray light curve of PKS~2332$-$017 
in 0.1--500\,GeV (upper), and its optical CRTS~V band and ZTF $zg$ and 
$zr$ band light curves (middle) and MIR WISE light curves (bottom). {\it Right:}
Zoomed-in $\gamma$-ray (upper) and optical/MIR (bottom) light curves in 
MJD~54683--58000. In both $\gamma$-ray light curve panels, the downward arrows 
are 
the 95\% C.L. flux upper limits, red histograms indicate the TS values of 
the data points, and the gray regions mark the flare defined by the BB 
algorithm.  The dashed red lines mark the arrival time of IC-130127A. 
\label{fig:IC130127_lc}}
\end{figure*}

\subsection{{\it Fermi} $\gamma$-ray Data and Source Model}\label{fermi}

High-energy $\gamma$-ray data from the Large Area Telescope (LAT) onboard
{\it the Fermi Gamma-ray Space Telescope (Fermi)} were used by us for studying 
the two blazar targets. Photon events in the energy range of 0.1--500 GeV 
(evclass=128 and evtype=3) from the updated Fermi Pass 8 database in a time 
range of from 2008-08-04 15:43:36 (UTC) to 2024-04-18 00:05:53 (UTC) were 
obtained. For a target, the region of interest (RoI) was set to be 
20\degr $\times$ 20\degr\  with the center at the position of this target. 
We excluded the events with zenith angles $>$ 90\degr\ to avoid 
the contamination from the Earth limb, and the expression 
DATA\_QUAL $>$ 0 \&\& LAT\_CONFIG = 1 was used for selecting good time-interval 
events. In our analysis, the package Fermitools-2.2.0 and the instrumental 
response function P8R3\_SOURCE\_V3 were used.

For each target, a source model was constructed based on the latest
(Data Release 4, DR4) {\it Fermi}
Gamma-ray LAT (FGL) 14-year source catalog, 4FGL-DR4 \citep{Ballet+23}. 
All sources in the catalog
within 25\degr\ of a target were included. The spectral models in 4FGL-DR4 
for the sources, including our targets, were adopted. 
We set the spectral indices and normalizations of the sources within 
5\degr\ of each target as free parameters and fixed the other parameters 
at the catalog values. We also included the extragalactic diffuse emission 
and the Galactic diffuse emission components, which were the spectral files 
iso\_P8R3\_SOURCE\_V3\_v1.txt and gll\_iem\_v07.fits respectively. 
We always set the normalizations of these two components as free parameters 
in our analysis.

\section{Data Analysis and Results}\label{ana}

\subsection{\rm IC-130127A}
\subsubsection{Positional analysis}

IC-130127A was a GFU Gold type event \citep{Abbasi+23}
with an energy of 235\,TeV and a 
signalness of $\sim$ 61\% (the probability of astrophysical origin).
It was detected on 2013 January 27 06:43:11.01 UT (MJD 56319.28), with
a position of R.A. = 352$\fdg$97$_{-1\fdg01}^{+1\fdg32}$, 
Decl. = $-$1$\fdg$98$_{-0\fdg90}^{+0\fdg97}$  (equinox J2000.0, 90\% uncertainty)
in IceCat-1. In 4FGL-DR4, there are two $\gamma$-ray sources within the 
positional uncertainty region (see left panel of Figure \ref{fig:maps}), which
are 4FGL~J2335.4$-$0128 (or PKS 2332$-$017) and 4FGL~J2333.4$-$0133 
(or PKS B2330$-$017). Both are FSRQs. We checked their 1-yr binned
$\gamma$-ray light-curves provided by 4FGL-DR4\footnote{\url{https://fermi.gsfc.nasa.gov/ssc/data/access/lat/14yr_catalog/}}, 
and found that PKS 2332$-$017 had a significant flare at the arrival time 
of IC-130127A while PKS B2330$-$017 did not have significant flux variations 
at the time.

\subsubsection{Light-curve analysis for PKS~2332$-$017}
can
PKS~2332$-$017 was modeled as a point source with a Log-Parabola (LP) spectrum,
$dN/dE = N_0 (E/E_b)^{-[\alpha +\beta\ln(E/E_b)]}$ in 4FGL-DR4. Setting this
spectral model, we performed the standard binned likelihood analysis to 
the whole data in 0.1--500\,GeV for PKS~2332$-$017.  We obtained 
$\alpha$ = 2.36$\pm$0.13, $\beta$ = 0.27$\pm$0.10, and photon flux 
F$_{\gamma}$ = (6.40$\pm$2.38)$\times$10$^{-9}$~ph\,cm$^{-2}$\,s$^{-1}$, with 
a test statistic (TS) value of 351 (detection significance $\approx$$\sqrt{TS}$
$\approx 18.7\sigma$). These parameter values are consistent 
with those given in 4FGL-DR4, but the TS value is higher than that in
the catalog (TS = 124). The reason for the difference is that the source
recently had a flare (see left panel of Figure \ref{fig:IC130127_lc}).

We extracted the 0.1--500\,GeV $\gamma$-ray light curve of PKS 2332$-$017 by 
setting a 180-day time bin and performing the maximum likelihood analysis 
to the data of each time bin. In the extraction, only the normalization 
parameters of 
the sources within 5\degr\ of the target were set free and the other parameters
were fixed at the best-fit values obtained in the analysis of the whole data. 
For the obtained light-curve data points with TS $<$ 4, we calculated 
the 95\% confidence level (C.L.) upper limits and used the upper limits
instead.
As shown in the left panel of Figure~\ref{fig:IC130127_lc}, the source had
two flares and
the arrival time of the neutrino IC-130127A was temporally coincident with 
the first one.
Using the Bayesian block (BB) algorithm \citep{Scargle+13}, implemented 
through Python package 
Astropy\footnote{{https://docs.astropy.org/en/stable/api/\\ astropy.stats.bayesian\_blocks.html}}, 
we determined the time durations of the two flares,
the first being  $\sim$4\,yr (MJD~55582.66--57022.66, gray region in 
Figure~\ref{fig:IC130127_lc}) and the second being $\sim$3.5\,yr
(MJD 59542.66--60442.66). For the time ranges excluding the two flares, we 
defined them as the quiescent state of the source. 

We also obtained the optical light curves of the source from ZTF and 
CRTS and MIR ones from WISE. Focusing on the first flare
(right panel of Figure \ref{fig:IC130127_lc}),
it can be seen that the optical and 
MIR variations accompanied the $\gamma$-ray flare, and the brightening
variations were more than $\sim$1\,mag. 

\subsubsection{Spectral analysis for PKS~2332$-$017}

We performed the likelihood analysis to the data of the first flare and 
quiescent state. For the flare, we obtained $\alpha$ = 2.47$\pm$0.11, 
$\beta$ = 0.15$\pm$0.07, and 
F$_{\gamma}$ = (1.68$\pm$0.43)$\times$10$^{-8}$\,ph\,cm$^{-2}$\,s$^{-1}$ 
(TS = 230 or detection significance $\approx$ 15.2$\sigma$). For the quiescent
state, we obtained $\alpha$ = 2.97$\pm$0.54, $\beta$ = 0.93$\pm$0.53, and 
F$_{\gamma}$ = (1.23$\pm$0.89)$\times$10$^{-9}$ ph cm$^{-2}$ s$^{-1}$ 
(TS = 20 or detection significance $\approx$ 4.5$\sigma$). Given the low
detection significance and the resulting large 
parameter uncertainties in the quiescent state, we could not determine
any spectral changes between the flare and the quiescent state. 

Because of the detection of VHE emission in the neutrino-emitting flare of 
TXS~0506+056 and other evidence of hardening of emission during the 
the flares of the candidate neutrino blazars 
(see \citealt{Ji+24} and references therein), 
we checked high-energy $\geq$2\,GeV photons from PKS~2332$-$017.
By running {\tt gtsrcprob} on the data during the first flare in 
an 1\degr\ RoI, we found 10 photons of $\geq$90\% probabilities coming from
the source, with the highest energy being 13.7\,GeV (on MJD 56127). As 
the calculations rely on the model considered, when we changed to use
the model results from the whole data instead, we found four such photons 
in the flare (including that 13.7\,GeV one) and one in quiescence. Therefore, 
more high-energy photons were observed during the flare.

To obtain the $\gamma$-ray spectra of PKS~2332$-$017 in the first flare 
and the quiescent state, we performed the binned likelihood 
analysis to the respective data in 8 energy bins evenly divided in logarithm 
from 0.1 to 500\,GeV. In the analysis, the spectral normalizations of 
the sources 
in the source model within 5\degr\ of PKS~2332$-$017 were set free and
all other spectral parameters of the sources were fixed
at the values obtained in the above likelihood analysis
to the respective data of the flare and the quiescent state. 
For the spectra, the data points with TS $\geq$ 4 were kept.

\begin{figure*}[!ht]
\centering
\includegraphics[width=0.63\textwidth]{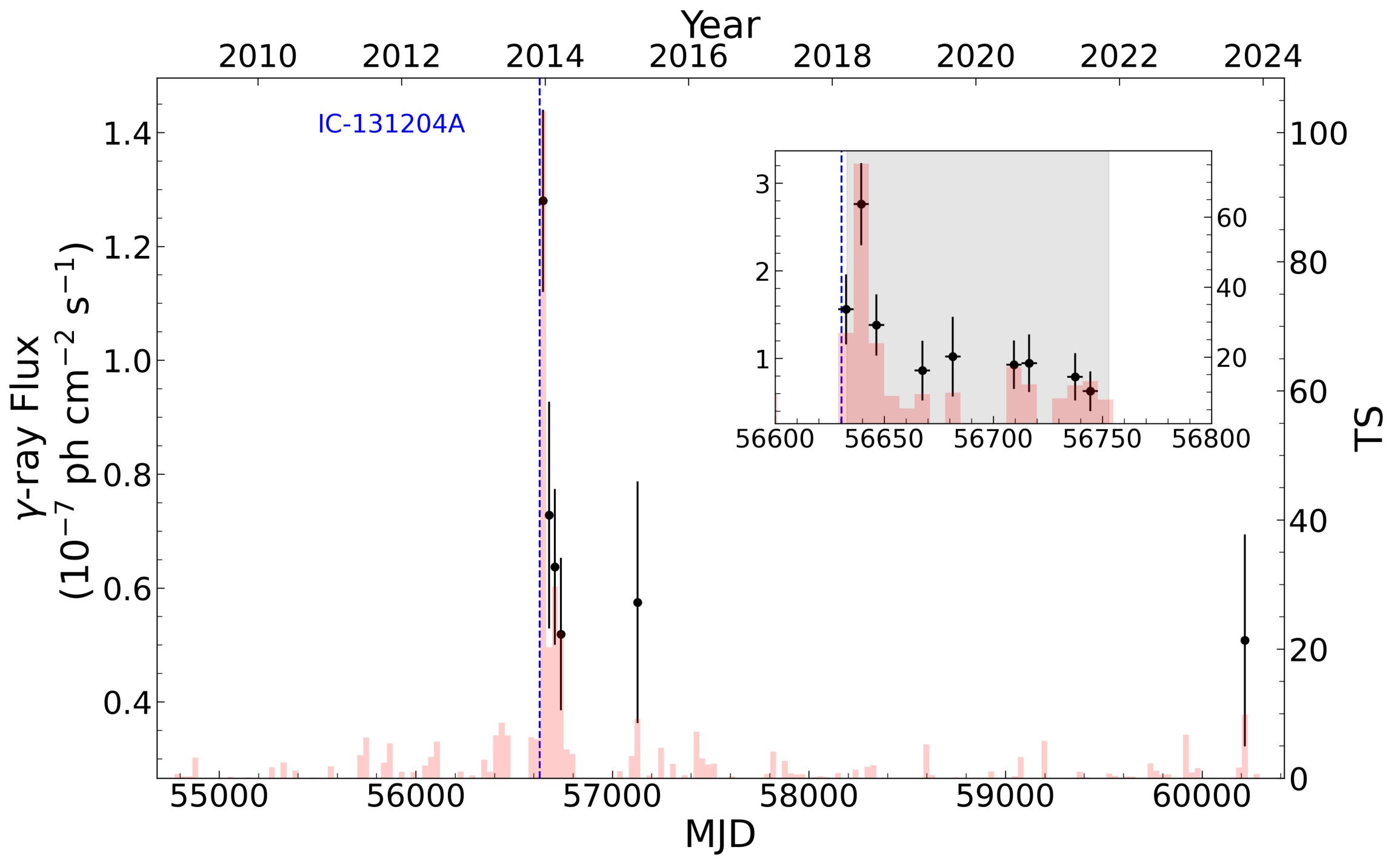}
\caption{30-day binned $\gamma$-ray light curve of PMN J1916$-$1519 
in 0.1--100\,GeV obtained from the {\it Fermi} LCR. The data points 
with TS $\geq$ 9 are kept in the light curve, 
and the red histograms indicate the TS values of the data points. 
The dashed blue line marks the arrival time of IC-131204A.
The sub-graph shows the zoomed-in 7-day binned light curve in MJD 56600--56800 
and the gray region marks the 4-month flare duration seen in the 30-day binned
light curve.
\label{fig:IC131204_lc}}
\end{figure*}

\subsection{\rm IC-131204A}
\subsubsection{Positional analysis}

IC-131204A, an EHE Gold type event with an energy of 259\,TeV and a signalness 
of $\sim$ 20\%, was detected on 2013 December 4 11:16:54.26 UT (MJD 56630.47). 
Its position was given to be R.A. = 288$\fdg$98$_{-0\fdg83}^{+1\fdg10}$, 
Decl. = $-$14$\fdg$21$_{-1\fdg31}^{+0\fdg77}$ 
(equinox J2000.0, 90\% uncertainty) 
in IceCat-1. There was only 4FGL J1916.7$-$1516 (or PMN~J1916$-$1519) within 
the uncertainty region (see right panel of Figure~\ref{fig:maps}). 
PMN J1916$-$1519 was 
classified as a blazar of uncertain type (BCU) in 4FGL-DR4.

\subsubsection{Light-curve analysis for PMN J1916$-$1519}

PMN J1916$-$1519 is being monitored by the {\it Fermi} LAT Light Curve 
Repository (LCR) program \citep{Abdollahi+23}. We obtained its 30-day binned 
$\gamma$-ray light curve from LCR, which is shown in 
Figure~\ref{fig:IC131204_lc}. Only data points with TS $>$ 9 are kept in 
the light curve. As can be seen, there was only a short flare, peaking at 
$\sim$ MJD~56648 and lasting $\sim$4 month in $\sim$MJD~56633--56753,
during the 16 year of the {\it Fermi} LAT data. 
The flare's time duration, simply the four-month bins that have
TS$>$20, is indicated by the gray region shown in 
the sub-graph of Figure~\ref{fig:IC131204_lc}.
The arrival time of IC-131204A approximately matches 
the flare, only $\sim$ 2.5 day before the first time bin (i.e., the peak)
of the flare. We also show the 7-day binned light curve from
LCR in Figure~\ref{fig:IC131204_lc}. The neutrino's arrival time 
is $\simeq$9\,day
before the peak of this finer light curve, and within the first time
bin (MJD~56629--56636) whose TS $\simeq$ 27.
We checked the optical and MIR light curves of 
PMN J1916$-$1519 in the CRTS, ZTF, and NEOWISE database (and also in
the other databases). Unfortunately, 
there were no optical and MIR measurements during the flare.

For PMN J1916$-$1519, 
we defined the time ranges excluding the flare as its quiescent state,
during which it was nearly not detectable.

\subsubsection{Spectral analysis for PMN J1916$-$1519}

PMN J1916$-$1519 was modeled as a point source with a LP spectrum in 4FGL-DR4. 
We performed the standard binned likelihood analysis to the whole data in 
0.1--500 GeV for PMN J1916$-$1519. We obtained $\alpha$ = 2.77$\pm$0.16, 
$\beta$ = 0.36$\pm$0.14, and 
F$_{\gamma}$ = (7.45$\pm$2.28)$\times$10$^{-9}$\,ph\,cm$^{-2}$\,s$^{-1}$, 
with a TS value of 102 (detection significance $\approx$ 10.1$\sigma$). 
The values are consistent with those given in 4FGL-DR4.
We then performed the likelihood analysis to the data of the flare and 
quiescent state. For the first, we obtained $\alpha$ = 2.58$\pm$0.10, 
$\beta$ = 0.16$\pm$0.08, and 
F$_{\gamma}$ = (3.76$\pm$0.63)$\times$10$^{-8}$\,ph\,cm$^{-2}$\,s$^{-1}$ 
(TS = 213 or detection significance $\approx$ 14.6$\sigma$). For 
the latter, we obtained $\alpha$ = 2.99$\pm$0.61, $\beta$ = 1.25$\pm$0.40 
and F$_{\gamma}$ = (1.72$\pm$0.70)$\times$10$^{-9}$\,ph\,cm$^{-2}$\,s$^{-1}$ 
(TS = 30 or detection significance $\approx$ 5.5$\sigma$). The 
results suggest that the emission during the flare is harder than that 
during the quiescent state, but still the uncertainties in the latter are
large, not allowing us to draw a certain conclusion. Running {\tt gtsrcprob}, 
we found four $\geq$2\,GeV photons of $\geq$90\% probabilities coming from 
the source, with the highest energy being
9.8\,GeV (on MJD 56742), during the flare, where the model of the flare was
used. When we changed to use the model from the whole data, no such photons
were found.

We obtained the $\gamma$-ray spectra of PMN J1916$-$1519 in the flare
and in the quiescent state by performing the binned likelihood analysis to the
respective data in 8 energy bins evenly divided in logarithm from 0.1 to 
500\,GeV. In this analysis, we set the spectral normalizations of the sources 
in the source model within 5\degr\ of the target as free parameters and
fixed the other spectral parameters 
at the values obtained in the above likelihood analysis
to the respective data of the flare and the quiescent state. 
Only data points with TS $\geq$ 4 in the spectra were kept.

\begin{figure*}[!ht]
\centering
\includegraphics[width=0.48\textwidth]{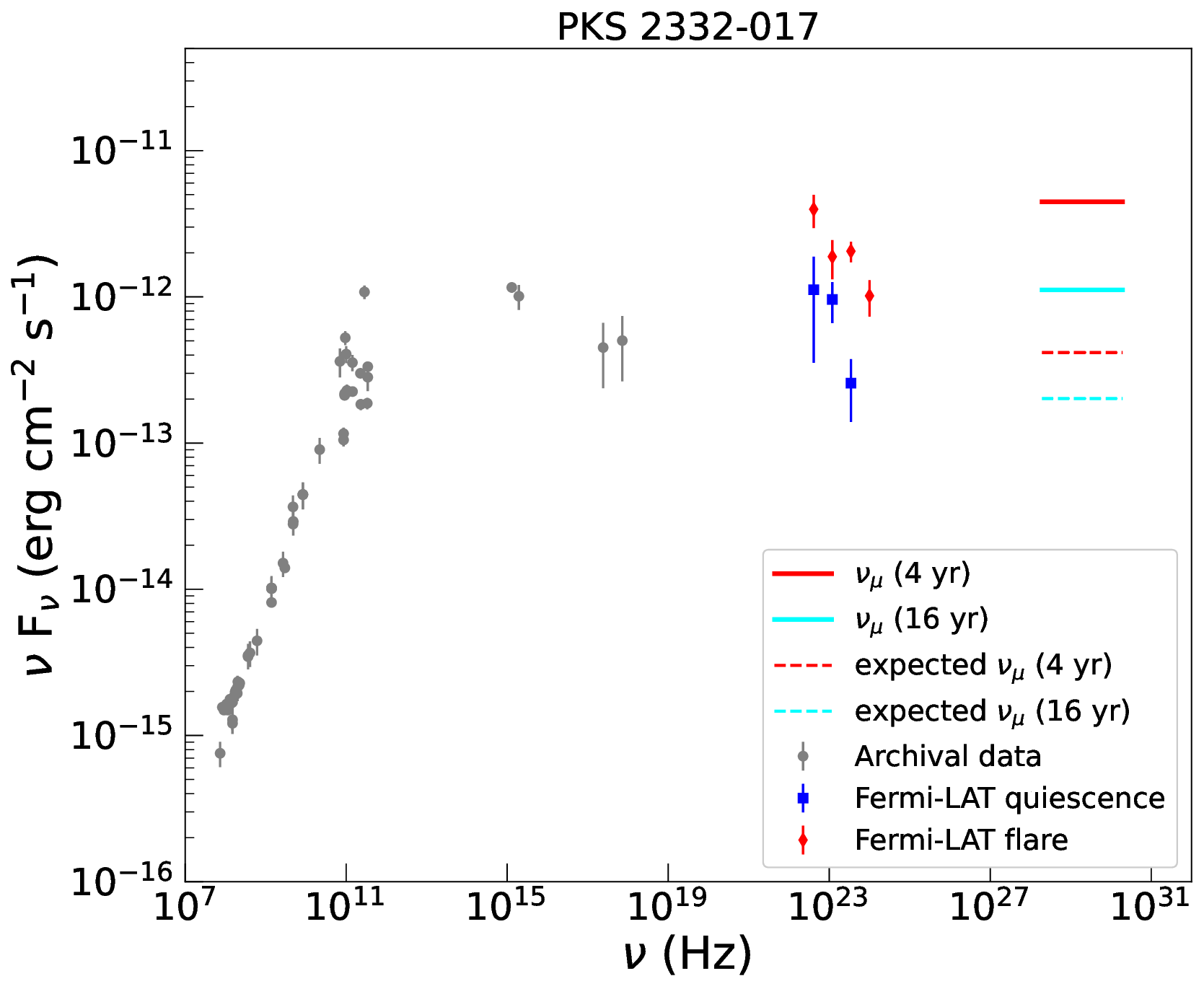}
\includegraphics[width=0.48\textwidth]{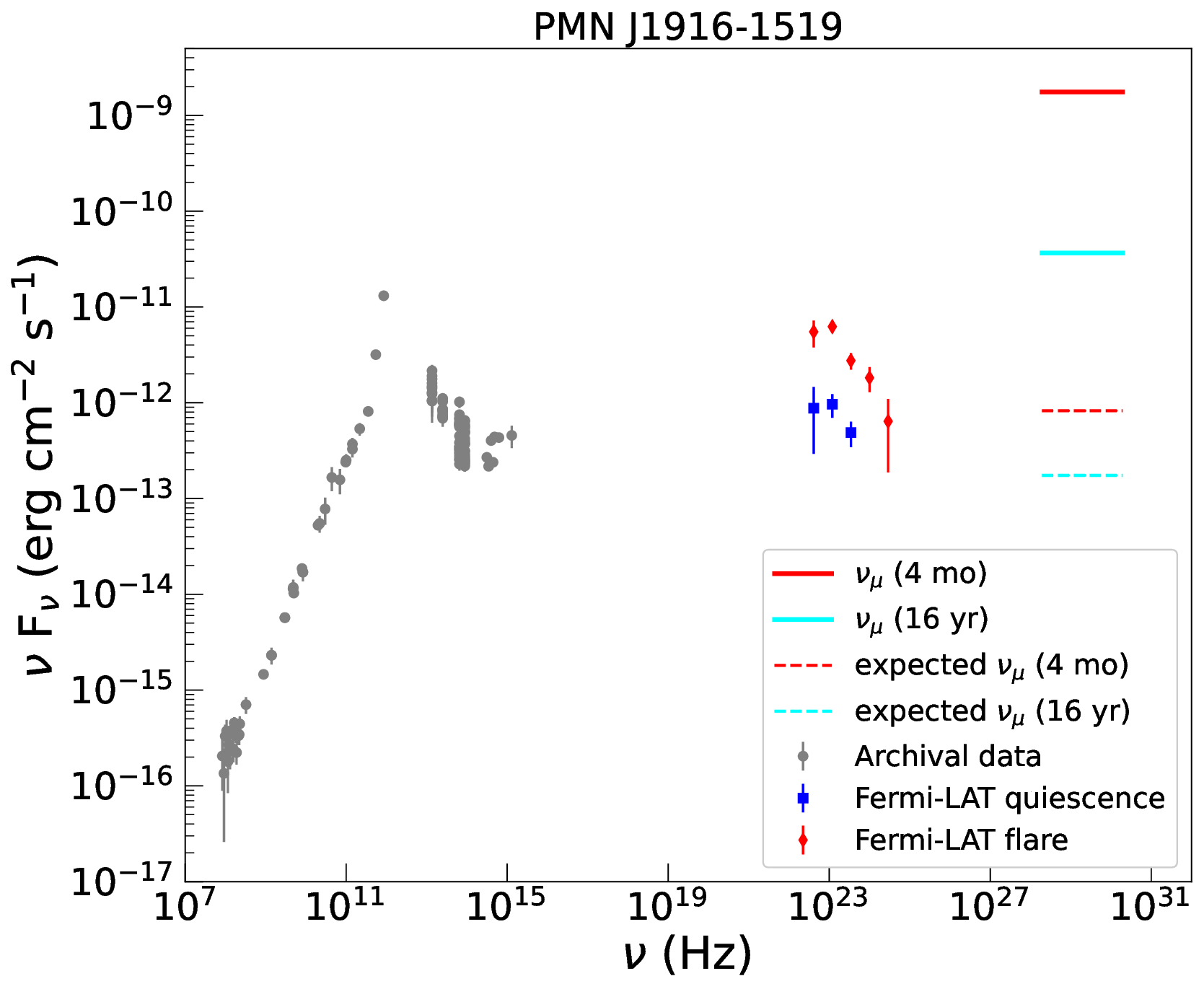}
\caption{SEDs of PKS 2332$-$017 ({\it left}) and PMN J1916$-$1519 ({\it right}).
The archival data points (gray) are from Firmamento, and the $\gamma$-ray
spectral data points we obtained are shown as red dots (flare) and 
blue squares (quiescence). The neutrino fluxes estimated from the detection
	of one neutrino with IceCube during the flare
and the whole {\it Fermi} LAT data for each source are plotted
as the red and cyan line respectively, while the neutrino fluxes estimated
	from the observed $\gamma$-ray luminosities are respectively shown
	as the dashed lines (see Section~\ref{dis} for details).
\label{fig:sed}}
\end{figure*}

\section{Discussion}\label{dis}

We went through the Gold events in IceCat-1, and found two blazars, 
PKS~2332$-$017 and PMN~J1916$-$1519, as the possible neutrino sources, in
addition to the previously reported neutrino blazars. The two sources 
are in positional coincidence respectively with IC-130127A and IC-131204A 
and had
a flare temporally coincident with the respective arrival times of 
the two neutrinos.
For PKS~2332$-$017, significant flux variations at optical and MIR bands
accompanying the flare were observed, which were also seen in nearly 
all previously reported neutrino 
blazars. For PMN~J1916$-$1519, due to the lack of multiband flux 
measurements over the 4-month flare, we do not have the information.
Although it is not possible to determine the emission hardening during
the flares for the two sources because of their weak emission during the
quiescent state, we have found high-energy photons, with energy as high as
$\sim$10\,GeV,
during the flares, which may be considered as evidence for emission hardening
\citep{Ji+24}.

PKS~2332$-$017 is an FSRQ at redshift $z$ = 1.18 \citep{Wills+78}. It is
classified as an LSP in the Data Release 3 (DR3) of the Fourth Catalog 
of Active Galactic Nuclei detected by the {\it Fermi} LAT
(4LAC-DR3; \citealt{Ajello+22}), with the  peak frequency 
$\nu^{\rm syn}_{\rm pk}\approx 2.52\times 10^{12}$\,Hz 
(in 4LAC-DR3) or $\approx 6.31^{+6.29}_{-3.15} \times 10^{12}$\,Hz given
by Firmamento\footnote{\url{https://firmamento.hosting.nyu.edu/data_access}} \citep{Firmamento}.
It is thus similar to several other reported neutrino blazars, PKS~B1424$-$418, 
GB6~J2113+1121, and NVSS~J171822+423948, which are all LSP FSRQs. 
Its long-term $\gamma$-ray luminosity and spectral index $\alpha$ were 
$3.7 \times 10^{46}$\,erg\,s$^{-1}$ and 2.36, and during the neutrino-arriving
flare (MJD 55582.66--57022.66), the values were 
$7.7 \times 10^{46}$\,erg\,s$^{-1}$ and 2.47, respectively. These values are 
typical for an FSRQ. Its SED is shown in
the left panel of Figure~\ref{fig:sed}.

PMN~J1916$-$1519 only has a photometric redshift $z$=0.968 \citep{Foschini+22}.
It is also an LSP blazar, as its 
$\nu^{\rm syn}_{\rm pk}\approx 6.24\times 10^{12}$\,Hz (in
4LAC-DR3) or $\approx 6.31^{+3.69}_{-2.33} \times 10^{12}$\,Hz given
by Firmamento.
Assuming the redshift, the $\gamma$-ray luminosities are estimated to 
be $2.0\times 10^{46}$\,erg~s$^{-1}$ in its long-term 16 year of {\it Fermi} 
LAT data and $9.4 \times 10^{46}$\,erg~s$^{-1}$ in its 4-month flare.
BL Lacs and FSRQs can be relatively well separated in the 
plot of $\gamma$-ray luminosity versus photon index 
(e.g., \citealt{Chen18}). Following the criterion line to separate BL Lacs 
and FSRQs from \citealt{Chen18}, PMN~J1916$-$1519 may be classified as an
FSRQ. Its SED is shown in the right panel of Figure~\ref{fig:sed}.

The $\gamma$-ray flare of PMN J1916$-$1519 in temporal coincidence with
IC-131204A was short, only $\sim$4 month.
Such short-flare cases were seen
before in other neutrino blazars. PKS~0735+178 had a $\sim$3-week 
$\gamma$-ray flare temporally coincident with the arrival time of 
IC-211208A \citep{sah+23}, GB6 J1040+0617 had a $\sim$3-month $\gamma$-ray 
flare (IC-141209A; \citealt{gar+19}), MG3 J225517+2409 showed a $\sim$5\,month
one (IC-100608A; \citealt{fra+20}), and TXS 0506+056 had a $\sim$ 6-month 
one (IC-170922A; \citealt{txs0506a}). PMN J1916$-$1519 would be unique as
it only had such a short flare over 16 year of the {\it Fermi} LAT observation
time.
According to the simulation results presented in \citealt{Capel+22}, 
the chance coincidence probabilities between $\gamma$-ray flares of blazars 
and neutrino alerts are high, maybe $\sim$3.8\%--12.7\%, but when certain 
source flux and flare variation amplitude are considered,
the probability could be lowered to $<$0.1\% (i.e., $>$3$\sigma$ 
significance for the source association). PMN~J1916$-$1519 could be such an 
example as its flare variation amplitude reached $\sim$5.0.

\begin{figure}[!ht]
\centering
\includegraphics[width=0.45\textwidth]{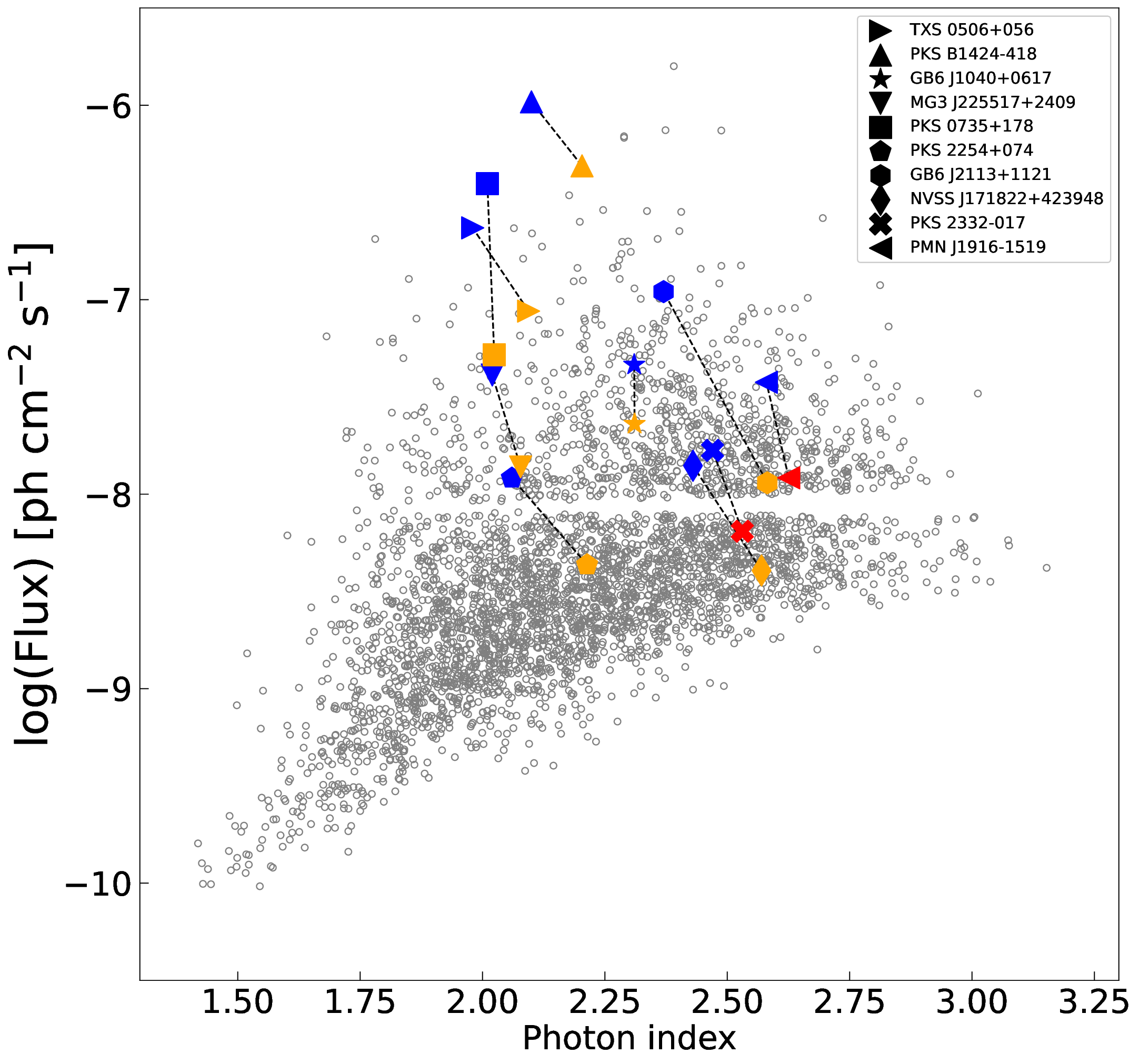}
	\caption{Flux (0.1--100\,GeV) and photon index of $\gamma$-ray blazars,
	with the values adopted from 4LAC-DR3. The previously reported
	(candidate) neutrino-emitting blazars are marked in yellow (the 
	long-term values) and blue (the flare values), while 
	the two sources in this work are marked in red (their long-term values)
	and blue (the flare values).
\label{fig:comp}}
\end{figure}

The redshift and $\gamma$-ray luminosity of PKS~2332$-$017 are comparable 
with the values of previously identified neutrino blazars. For example,
their redshift range is 0.19--2.68 and luminosity range is 
$\sim$ 10$^{44}$--10$^{48}$ erg s$^{-1}$ \citep{Ji+24}. 
The neutrino flux may be estimated from the observed $\gamma$-ray flux 
(see \citealt{gio+20} and \citealt{jiang+24}), using the relationship between 
them given by \citep{Murase+18},
\begin{equation}\label{ppi}
	\begin{split}
    {\epsilon_\nu}{L_{\epsilon_{\nu}}} \approx \frac{6\left(1+Y_{IC}\right)}{5}{\epsilon_\gamma}{L_{\epsilon_{\gamma}}}\vert_{\epsilon_{\text{syn}}^{p\pi}} \\
    \approx 8 \times 10^{44} \, \text{erg s}^{-1} \left(\frac{{\epsilon_\gamma}{L_{\epsilon_{\gamma}}}\vert_{\epsilon_{\text{syn}}^{p\pi}}}{7 \times 10^{44}}\right).
	\end{split}
\end{equation}
Here, $Y_{IC}$ represents the Compton-Y parameter, expected typically 
$Y_{IC}$ $\leq 1$ \citep{Murase+18}, and the
$\gamma$-ray luminosity term 
$\epsilon_{\gamma}L_{\epsilon_{\gamma}} \sim$ \Lagr$_{\gamma}/\ln(500\,{\rm GeV}/100\,{\rm MeV})$ $\approx$ $9.1\times 10^{45}$~erg~s$^{-1}$ (for 4 yr) and 
$4.4\times 10^{45}$\,erg~s$^{-1}$ (for 16 yr;
see \citealt{gio+20} for details); 
\Lagr$_{\gamma}$ is the observed $\gamma$-ray luminosity for the source. 
Then the average muon neutrino luminosities 
(i.e., $1/3{\epsilon_\nu}{L_{\epsilon_{\nu}}}$) would be 
$3.5\times 10^{45}$\,erg~s$^{-1}$ (4 yr) and 
$1.7\times 10^{45}$\,erg~s$^{-1}$ (16 yr). 
On the other hand, the number of muon (and antimuon) neutrinos, 
$N_{\nu_{\mu}}$, detected by IceCube during time interval $\Delta{T}$ at 
declination $\delta$ may be estimated from
\begin{equation}\label{number}
     {N_{\nu_{\mu}}}  = \int_{\epsilon_{\nu_{\mu, \text{min}}}}^{\epsilon_{\nu_{\mu, \text{max}}}} {A_{\text{eff}}(\epsilon_{\nu_{\mu}}, \delta) \phi_{\nu_{\mu}} \Delta{T}} \, d\epsilon_{\nu_{\mu}},
\end{equation}
where $\epsilon_{\nu_{\mu, \text{min}}}$ and $\epsilon_{\nu_{\mu, \text{max}}}$
are the minimum and maximum neutrino energy, respectively, and
$\phi_{\nu_{\mu}}$ is the differential energy flux of muon neutrinos. 
Using the muon neutrino flux estimated from Eq.~\ref{ppi} above and
assuming a power-law neutrino energy spectrum with an index of $-$2 
from 80 TeV to 8 PeV \citep{Oikonomou+21}, we find from Eq.~\ref{number}
the expected numbers of neutrinos to be 0.09 ($\Delta{T}=$ 4\,yr) and 
0.18 ($\Delta{T}=$ 16\,yr) for 
the effective area $A_{\rm eff} \approx 23$\,m$^{2}$ (GFU\_Gold, 
see \citealt{Abbasi+23}). In other words, the Poisson probabilities to detect 
one neutrino are $\sim$0.08 and $\sim$0.15, respectively. Alternatively, we may
estimate the neutrino flux from Eq.~\ref{number} by setting 
$N_{\nu_{\mu}} = 1$ (given the neutrino event), and 
the resulting energy fluxes for two $\Delta T$ values
are shown in Figure~\ref{fig:sed}.

Similarly for PMN~J1916$-$1519, where we used the photometric redshift,
we estimated its $\gamma$-ray luminosity term 
$\epsilon_{\gamma}L_{\epsilon_{\gamma}} \approx$ $1.1 \times 10^{46}$\,erg~s$^{-1}$ (for 4 month) and $2.3\times 10^{45}$\,erg~s$^{-1}$ (for 16 yr),
the average muon neutrino luminosities $\sim 4.2\times 10^{45}$\,erg~s$^{-1}$ 
(4 month) and $\sim 8.9\times 10^{44}$\,erg~s$^{-1}$ (16 yr). 
The expected numbers of neutrinos 
would be $\approx 5\times 10^{-4}$ (for 4 month) and $\approx 5\times 10^{-3}$
(for 16 yr), suggesting 
low neutrino detection probabilities of $<$1\% for PMN~J1916$-$1519. 
The low detection probabilities, including those estimated in 
3HSP~J095507.9+355101 \citep{gio+20}, NVSS J171822+423948 \citep{jiang+24},
and PKS 2254+074 \citep{Ji+24}, are actually consistent with the results
from studies of SEDs with one-zone lepto-hadronic models
(e.g., \citealt{Keivani+18,Petropoulou+20a}). The extremely low probability
in the flare case of PMN~J1916$-$1519 was also seen in the SED modeling of
3HSP~J095507.9+355101, as \citet{Petropoulou+20b} found a neutrino-detection
probability of 0.06\% for the X-ray flare of the blazar.

Finally, the plot of 0.1--100\,GeV flux versus photon index of
$\gamma$-ray blazars (Figure~\ref{fig:comp}), which was previously shown 
in \citet{liao+22}, is updated by us. Five more sources, 
PKS~0735+178 \citep{sah+23}, NVSS J171822+423948 \citep{jiang+24}, and
PKS~2254+074 \citep{Ji+24}, plus two sources in this work, are added in 
the plot (but we do not include 3HSP J095507.9+355101 since its temporal 
coincidence was between an X-ray flare and a neutrino).
Now there are 10 flaring blazar candidates in total. Their average fluxes, 
given by the long-term {\it Fermi} LAT data, are in a wide range of 
from $4\times 10^{-9}$ to $5\times 10^{-7}$\,ph\,s$^{-1}$\,cm$^{-2}$. 
However, as seen in the 
two sources of this work, one/two flares can dominate the fluxes. 
In other words, the emission activity of such blazars, with PMN~J1916$-$1519 
as an extreme case, is mainly contributed by flaring events.
Thus as already shown by the case of TXS~0506+056, it is very likely
that the neutrino emission in blazars is closely related to the
flares. This is also supported, for example, by the blazar population studies 
\citep{Capel+22}.
 
\begin{acknowledgments}

This work was based on observations obtained with the Samuel Oschin Telescope 
48-inch and the 60-inch Telescope at the Palomar Observatory as part of the 
Zwicky Transient Facility project. ZTF is supported by the National Science
Foundation under Grant No. AST-2034437 and a collaboration including Caltech, 
IPAC, the Weizmann Institute for Science, the Oskar Klein Center at Stockholm 
University, the University of Maryland, Deutsches Elektronen-Synchrotron
and Humboldt University, the TANGO Consortium of Taiwan, the University of 
Wisconsin at Milwaukee, Trinity College Dublin, Lawrence Livermore National 
Laboratories, and IN2P3, France. Operations are conducted by COO, IPAC, and UW.

This work made use of data products from the Wide-field Infrared Survey 
Explorer, which is a joint project of the University of California, 
Los Angeles, and the Jet Propulsion Laboratory/California Institute of 
Technology, funded by the National Aeronautics and Space Administration.

This research is supported by the Basic Research Program of Yunnan Province 
(No. 202201AS070005), the National Natural Science Foundation of China 
(12273033), and the Original Innovation Program of the Chinese Academy of 
Sciences (E085021002). S.J. and D.Z. acknowledge the support of the science 
research program for graduate students of Yunnan University (KC-23234629).

\end{acknowledgments}

\bibliographystyle{aasjournal}
\bibliography{neu}{}

\end{document}